\documentclass[superscriptaddress,twocolumn, nature]{article}
\usepackage{graphicx}
\usepackage{caption}
\usepackage{subcaption}
\usepackage[superscript,biblabel]{cite}
\usepackage{authblk}
\usepackage{color}

\definecolor{fm}{rgb}{0.64, 0.0, 0.0}

\definecolor{ss_color}{rgb}{0.1,0.7,0.1}

\definecolor{st_color}{rgb}{0.5,0.5,0}

\definecolor{lightcyan}{rgb}{0.88, 1.0, 1.0}
\definecolor{lightorange}{rgb}{0.93,0.93,0.93}
\usepackage{amssymb}
\usepackage{tcolorbox}
\usepackage{multicol}
\usepackage{fancyhdr}
\usepackage{color, colortbl}
\usepackage{titlesec}
\usepackage{multirow}
\usepackage{stackengine}

\begin{document}

\titleformat{\subsubsection}[runin]{\normalfont\normalsize\bfseries}{\thesubsubsection}{5pt}{}[.]
 \title{Experimental vulnerability analysis of QKD based on attack ratings}
\author[1]{Rupesh Kumar}
\author[2]{Francesco Mazzoncini}
\author[3]{Hao Qin}
\author[2]{Romain All\'eaume}
\affil[1]{Quantum Communications Hub, Department of Physics, University of York, Heslington, YO10 5DD, United Kingdom.}
\affil[2]{ T\'el\'ecom Paris-LTCI, Institut Polytechnique de Paris, 19 Place Marguerite Perey, 91120 Palaiseau, France.}
\affil[3]{CAS Quantum Network Co., Ltd No. 99 Xiupu Rd, Pudong New District, Shanghai 201315, China.}
\twocolumn[
  \begin{@twocolumnfalse}
    \maketitle
\begin{abstract}

Inspired by the methodology used for classical cryptographic hardware, we consider the use of attack ratings in the context of QKD security evaluation.  To illustrate the relevance of this approach, we conduct an experimental vulnerability assessment  of CV-QKD against saturation attacks, for two different attack strategies. 
The first strategy relies on inducing detector saturation by performing a large coherent displacement. This strategy is experimentally challenging and  therefore translates into a high attack rating. We also propose and experimentally demonstrate a second attack strategy that simply consists in saturating the detector with an external laser. The low rating we obtain indicates that this attack constitutes a primary threat for practical CV-QKD systems. 
These results highlight the benefits of combining theoretical security considerations with vulnerability analysis based on attack ratings, in order to guide the design and engineering of practical QKD systems towards the highest possible security standards.
\vspace*{1cm}
\end{abstract}
  \end{@twocolumnfalse}
]

\section*{Introduction} 

Quantum Key Distribution (QKD) is a beautiful idea \cite{Bennett1984, Gisin2002}{}, that allows two legitimate users, the sender Alice and the receiver Bob, to establish a secret key with information-theoretic security (ITS) even against a quantum attacker, Eve. 

 QKD systems with increasing performances and reliability have been engineered over the past 25 years\cite{NPJQI2016, pirandola2019advances, xu2020secure}{}. 
 This has placed QKD among the quantum technologies with the highest maturity level. It has also paved the way for large-scale deployments and for the effective demonstration of QKD in relevant real-world  application contexts, requiring the ability to provide long-term security for data at flight or at rest\cite{sasaki2018quantum}{},  such as private clouds or critical infrastructures for government, defense or health data management \cite{lewis2019secure}{}.
A necessary condition for the industrial take-off of QKD will, however, reside not only in the ability to engineer cost-effective QKD systems, but also in the capacity to provide solid guarantees regarding their security.

Theoretical security proofs \cite{renner2008security,  scarani2009security, lo2014secure}  constitute a strong conceptual framework to capture the security properties of QKD protocols, based on a model.
QKD implementations may, however, not fully comply with the model used in the security proof, leading to security vulnerabilities and the possibility to launch side-channel attacks \cite{xu2020secure}{}.
Optimizing the real-world security of a QKD system hence requires to consider not only the {\it theoretical security} of the QKD protocol, but also the {\it practical security} related to its implementation. 
%
As a matter of fact, engineering constraints may impose stringent limits to the security level that a QKD system can reach. Such constraints can indeed lead to significant deviations between the theoretical security level that could be expected with an idealized implementation and the security level that can be reached in practice by the real QKD system.

As system design and security evaluation are in practice almost always  limited by resources, attacks that are  easier to implement should be prioritized, as they represent the greatest threats. For instance,  some attacks on QKD can be realized with a  relatively simple procedure and inexpensive hardware, such as detector blinding attack\cite{Lydersen2010} that has even been demonstrated on a live QKD connection \cite{Gerhardt2011}{}. Some other attacks, on the other hand, such as the photon number splitting attack \cite{brassard2000limitations} and more generally collective and coherent attacks on QKD \cite{renner2008security}{}, have played a fundamental role in our understanding of QKD theory. Yet, their implementation requires the ability to store and retrieve single photons  from a quantum memory, potentially over ms or larger timescales, which is currently out of reach, given the limitations of existing quantum memory technology \cite{lvovsky2009optical}{}.
Hence, to guarantee a very high security level for QKD, forward-looking methods and standards in quantum cryptography implementation security shall be adopted, following a methodology similar to the one used to certify the security of classical crypto-systems \cite{herrmann2002using}{}, called Common Criteria.

To demonstrate the relevance of this methodology for QKD, we consider here  continuous-variable (CV)-QKD  \cite{Grosshans2002,Grosshans2003} system and specific side channel attacks on its implementation.
In CV-QKD  a few attacks  have been proposed to exploit the security loopholes \cite{Haseler2008,Huang2013,Ma2013,Huang2014}{}, however most of them have been demonstrated off-line. Demonstrating an active attack on a live CV-QKD system is challenging. It requires to overcome several difficulties such as the technical complexity of the attack strategy itself, or of the optical phase recovery.
In this work, we demonstrate an active attack on a live CV-QKD system running the Gaussian modulated coherent state (GMCS)  protocol \cite{Grosshans2002}{}. Exploiting the non-linear response of the homodyne detector near its detection limit, an eavesdropper, Eve, can launch  an attack called Saturation Attack \cite{Qin2016,Qin2018}{}.

We have considered two practical methods to mount the saturation attack in CV-QKD. The first and most challenging  one is the coherent attack strategy \cite{Qin2016}{}, where Eve resends  coherent displaced signal \cite{Paris1996} to induce the detector saturation. 
Our second attack strategy  consists in the incoherent saturation attack\cite{Qin2018}{}, where we shine an independent laser towards Bob's coherent receiver.
The implementation of this attack is considerably simpler and it constitutes a dangerous threat to practical CV-QKD systems.
Inspired by the Common Criteria  Common Evaluation Methodology v3.1(CEM) \cite{CEM}{}, we introduce a metric called Attack Potential to QKD, and we evaluate the two aforementioned saturation attack strategies against this metric. 

Although the Attack Potential approach to a practical security evaluation is not new in the world of IT security,  its introduction in the context of QKD brings fresh perspectives. It has the ability to strengthen the security rationale associated with QKD system design and to accelerate  the evolution towards a QKD industry capable of manufacturing QKD devices with high security assurance.

\section*{Results}

%
%
%
%
%
%
%
%
%

\subsubsection*{Attack rating}
One crucial part of the complex methodology for this security evaluation is the process of identifying, classifying and prioritizing threats associated to vulnerabilities in QKD systems.  A comprehensive methodology offers  general guidance  and a metrics to rate the possible attacks against the assets. It also considers both the likelihood that a threat agent may successfully perform the attack  and the magnitude of the impact that this attack has on the assets.  In our rating procedure we shall focus on the likelihood of an attack, evaluating the total effort required to successfully mount the attack, called the Attack Potential: the higher the Attack Potential, the lower the chances of the attack being performed are.
To determine the Attack Potential we consider different factors (such as the type of equipment required); for each of them we assign a numerical value, the sum of them is the actual Attack Potential. In Table \ref{rating} we define the  semi-qualitative correspondance between rating and attack difficulty. Attack paths with an Attack Potential (AP) between $0$ and $10$ are for example rated as \textit{Basic}. Such attacks can be implemented with little effort and therefore constitute very serious threats. On the other hand, attacks with an extremely high Attack Potential, rated here \textit{Beyond High}, are extremely difficult to implement and therefore constitute less pressing threats.

\begin{table}[htb!]
\centering
\begin{tabular}{|c|c|}
\hline
\rowcolor{lightorange}
\textbf{Rating} & \textbf{AP Range} \\
\hline
\cellcolor{red!40}
Basic & $0-10$\\
\hline
\cellcolor{orange!40}
Moderate & $11-15$  \\
\hline
\cellcolor{yellow!40}
High & $16-19$\\
\hline
\cellcolor{lightcyan}
Beyond High &  $20-\infty$ \\
\hline
\end{tabular}
\caption{ Semi-qualitative scale for attack rating.
This scaling is adapted  with respect to the Common Evaluation Methodology  \cite{CEM}{}, taking into account the fact that we consider 4 out 5 factors in our analysis \cite{rating}{}. For a  more detailed discussion on the attack rating factors see Methods. }
\label{rating}
\end{table}

\subsubsection*{The saturation attack}
Saturation attack on CV-QKD consists in biasing the excess noise estimation by actively inducing the saturation of the homodyne detectors. This attack can be powerful\cite{Qin2016}{}: it can be combined with simple attack strategies by Eve (such as the intercept-resend attack \cite{Lodewyck2007a}) and lead to a full security breach. 

In a CV-QKD system that implements the GMCS protocol, Alice prepares coherent states of quadratures $\{X_A, P_A\}$, modulates each quadrature according to a Gaussian distribution of variance $V_A$ and centered on zero, and sends the modulated coherent state to Bob through the quantum channel. Bob  randomly measures one of the quadratures using a balanced homodyne detector. This results in quadrature measurements  $X_B$  and  $P_B$, with variance $V_B$. By correlating  sent and  measured  quadrature values on a fraction of their data, the users  estimate channel transmittance  $T$ and then the excess noise $\xi$. If these values  are within the limit for validating the security of the key, they proceed to key distillation on rest of the data, if not the QKD protocol aborts.

The balanced homodyne detector is a crucial part of a CV-QKD system. The linearity range of the homodyne detector response is in general characterized offline, as part of the detector calibration process. Assuming it is the same for $X$ and $P$ quadrature, we can designate this linearity range as a quadrature interval $\left[\alpha_1, \alpha_2 \right] (\alpha_1 < 0 < \alpha_2) $.

Balancing the homodyne detector prior to protocol run ensures that the mean of the homodyne output values remain close to zero, and therefore that the homodyne receiver is operated in its linear range, except if signals with very large quadrature values are received at Bob side. In case signals with  $X_{B} << \alpha_{1}$ (respectively $X_B >> \alpha_2$) are received, then the detector will saturate and output$ X_{B} = \alpha_{1} $ (respectively  $X_B = \alpha_2$).
 Close or beyond the detection limits, response of the detector becomes non-linear to the input signal quadrature, which will effectivement reduce the measured variance, compared to the actual quadrature variance of the received optical signals.

In the saturation attack, in order to reach the non-linear regime of the detector response,  Eve performs an intercept-resend attack and displaces  the mean of the Gaussian  quadrature modulation of the resent signals from  zero to a value $\Delta = \sqrt{\Delta_X^2 + \Delta_P ^2}$.  Without loss of generality, we can set $\Delta_X$ equal to  $\Delta_P$.  To perform the saturation attack Eve sets the amount of displacement  $\Delta$  such that  quadrature of the coherent state received by Bob  overpassed the linear range. As a consequence, the saturation affects the measured quadrature variance by Bob. Additionally, Eve may not only displace the quadrature value of the coherent state she resends to Bob, she may also apply some  amplification factor $G$ on the resent signal primarily  to  compensate the 3dB loss occurring from her heterodyne measurement during IR attack. 

Now, from the quadrature measurement data obtained from the saturated induced detector,  Alice and Bob estimate channel parameters  $T_{sat}$ and  $\xi_{sat}$ both are influenced by  displacement $\Delta$ and amplification factor $G$.  For a given value of Alice's quadrature modulation variance $V_A$, Eve can optimize $\Delta$ and $G$ such that excess noise $\xi_{sat}$ drops below  the null key noise threshold  and  eavesdropping remains undetected.
In such case, even though Eve has mounted an entanglement-breaking intercept-resend attack (which should lead the QKD protocol to abort due to a too high excess noise, generating no key) the attack is not detected, due to saturation, and Alice and Bob generate key, that is however insecure: this constitutes a characterized security break.

To characterize the attack, and in particular its impact on key rate, we have defined the following conditions.


%

    \begin{itemize}
    \item The attacker, Eve, performs the saturation attack: Intercept-resend attack combined with displacement.
    \item The channel transmission estimation is  unaffected ($T_{sat} = T$, where $T$ is the channel transmission in absence of attack).
    \item Alice and Bob obtain a positive key rate from their estimated parameter $T_{sat}$ and $\xi_{sat}$.
    \end{itemize}
  
\begin{figure}[htb!]
 \includegraphics[width=0.5\textwidth]{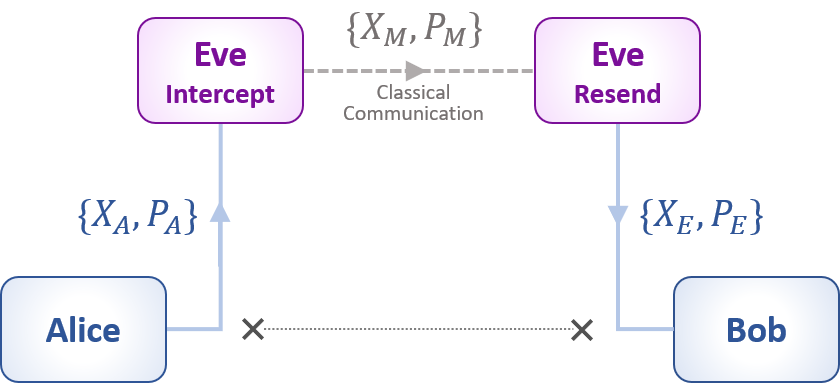}
                \caption{Scheme for saturation attack.  Eve$_{intercept}$  intercepts Alice's Gaussian modulated signal of  quadratures  $\{X_A,P_A \}$  and shares her measurement results $\{X_M,P_M\}$ through the classical channel to  Eve$_{resend}$. The resent and displaced signal of quadrature $\{X_E, P_E\}$ is measured by Bob homodyne detector.}
                \label{Setup_Basic}      
\end{figure}

The actual realization of the saturation attack comprises of two steps:  intercepting Alice's signal and resending a newly prepared signal to Bob with displacement $\Delta$ and gain $G$. We can consider that two cooperating eavesdroppers are involved in the attack:  Eve$_{intercept}$, located near  Alice intercepts the signals of  quadratures  $\{X_A,P_A \}$ and classically communicates the measurement results $\{X_M,P_M\}$ to Eve$_{resend}$- located near to Bob as shown in Figure \ref{Setup_Basic}.  Due to the technical restrictions imposed by the laboratory equipment, we experimentally demonstrate only the resend step of the attack and model the impact of the measurement associated with the intercept step. $\{X_M, P_M\}$  is deduced from $\{X_A,P_A \}$ by simulating a heterodyne measurement, i.e. 3 dB loss factor and also the addition of a random Gaussian noise of variance 2 shot noise \cite{Lodewyck2007a}.

\subsubsection*{Coherent attack strategy}

The  signal of quadrature $\{X_E, P_E\}$ is resent by Eve$_{resend}$, that we will from here onwards label as Eve. 
is experimentally generated, using a setup built around a Sagnac interferometer, represented on Figure \ref{Setup_Displaced}, and whose functioning is detailed in Methods.
The role of this set-up is to generate, knowing the in values $\{X_M,P_M\}$,  a displaced coherent state of quadrature $\{X_E, P_E\}$ that are correspond to the encoding of $\{X_M,P_M\}$  on a coherent states, to which is applied a coherent gain $\sqrt{G/2}$ in amplitude, and a controlled coherent displacement by a value  $\Delta = \Delta_X = \Delta_P$.
 The Sagnac loop offers a high phase stability which allows to precisely control $\Delta$ and therefore minimize the noise. Receiving the displaced coherent state $\{X_E, P_E\}$, Bob randomly measures one of the quadratures  with a balanced homodyne detector, cf Figure \ref{BalancedHD}, hence obtaining $X_B$ or $P_B$. Depending on the value of $\Delta$, this quadrature measurement will  be obtained in the linear or in the saturated regime.

\begin{figure}[htb!]
 \includegraphics[width=0.47\textwidth]{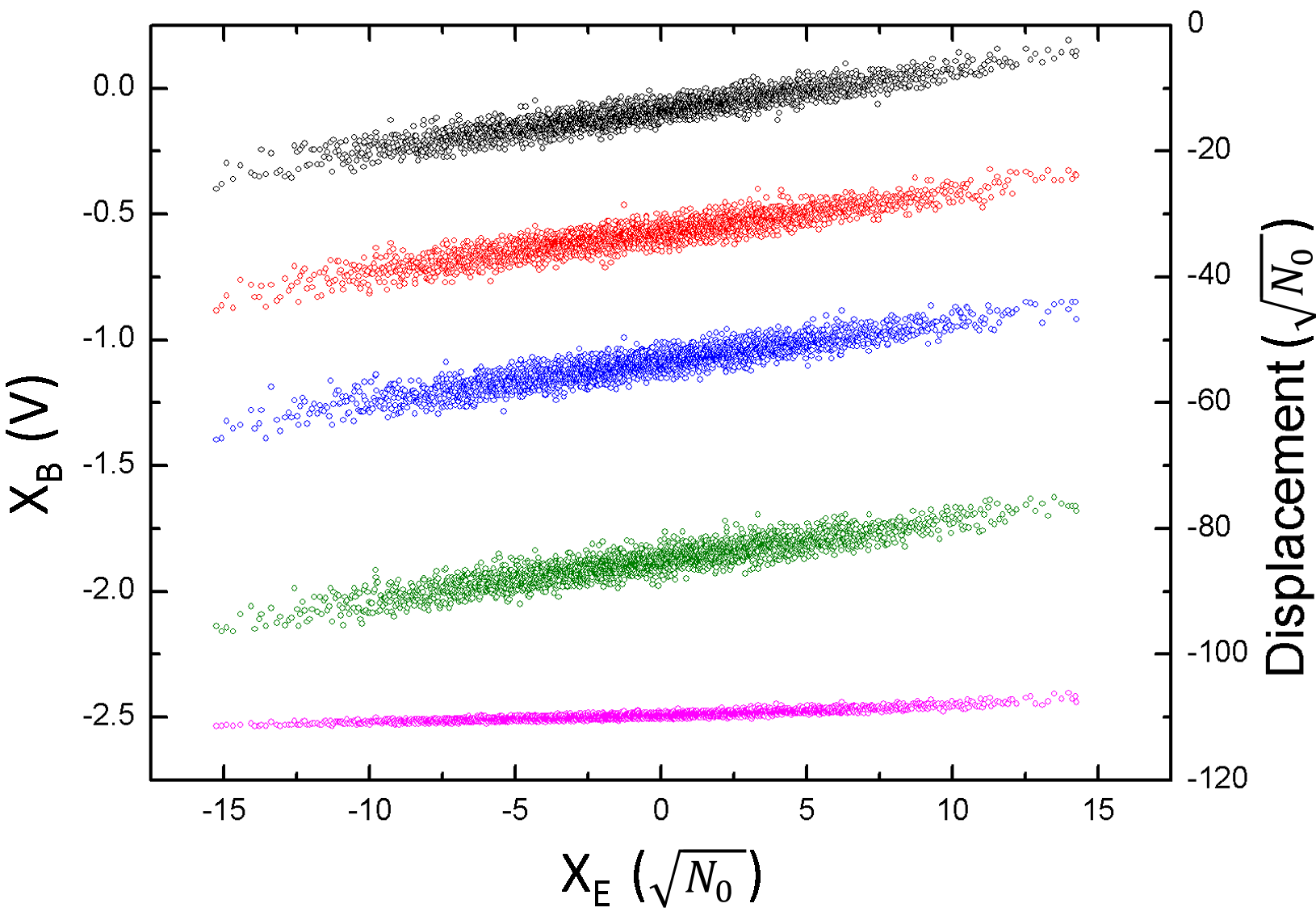}
                \caption{Response of homodyne output due to displacement. Given an input signal sent by Eve, with quadrature variance $Var(X_E)=22N_0$, it shows how the distribution of  Bob quadrature measurement results $X_B$ varies as a function  of displacement $\Delta$.}
                \label{Displacement}       
\end{figure}

Figure \ref{Displacement} shows the effect of the displacement $\Delta$ on Bob's experimental quadrature measurements. The mean value of the homodyne output $X_B$ can be shifted towards one of the detection limit $\alpha_1 = -2.5V$ of the detector, for a given displacement setting .  Selecting displacement angle to 225 degree would direct the shift towards the other limit of the linear range $\alpha_2 = 3.3V$. As can be seen on Figure \ref{Displacement}, when the detector is operated close to it linear range limit, then saturation occurs and quadrature variance reduces drastically.

The coherent displacement set-up demands an active feedback routine to compensate the relative phase drifts between the displaced signal and the local oscillator. Even though Sagnac loop provides a high stability, as illustrated by the level of control obtained on Figure \ref{Displacement}, we could not lower the residual quadrature noise due to imperfect phase drift compensation below the null key threshold. For example, considering that  2$\pi$ phase drift occurring in 1 second, a 500 $\mu$s  latency  in the feedback loop creates about 0.2 degrees of phase error. This in turn results in 0.23$\sqrt{N_0}$ fluctuations in homodyne output and generates excess noise of about $5 N_0$. This implies that the excess noise $\xi_{sat}$ is above the null key noise threshold value, and prevents the generation of key.  In other words, in the current setup, Alice and Bob would easily detect  attack based on coherent displacement.  Reducing the feedback latency 
such that phase drift remains negligible within the feedback intervals, could however bring this attacking strategy to meet 
the attack success conditions.

\subsubsection*{Incoherent attack strategy}
\begin{figure}[htb!]
 \includegraphics[width=0.5\textwidth]{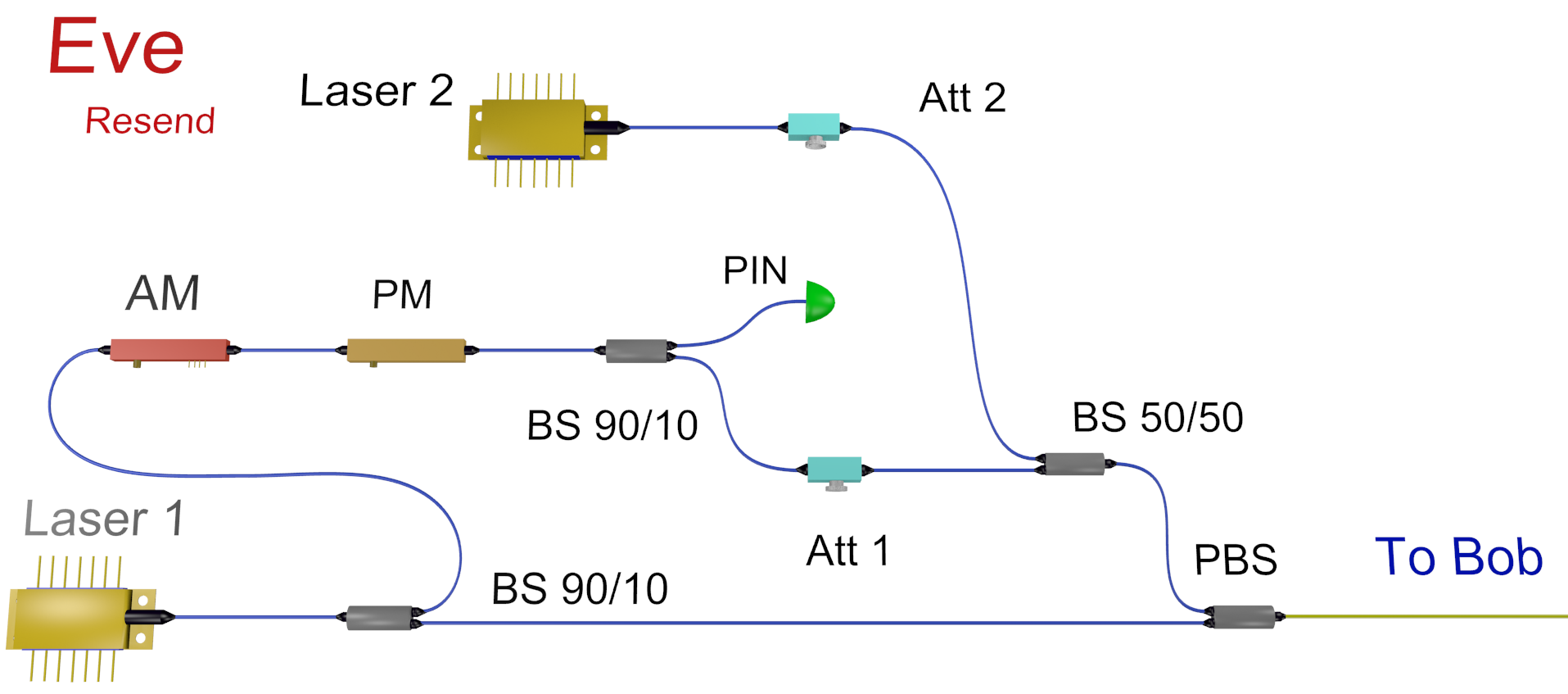}
                \caption{Setup for incoherent coherent attack strategy, relying on pulse injection from an external incoherent laser to induce saturation. AM: Amplitude Modulator, PM: phase Modulator, BS: BeamSplitter, PBS: Polarization BeamSplitter, Att: Variable Attenuator.}
                \label{Setup_Ext_Laser}      
\end{figure}

 In order to  overcome the implementation difficulties of the coherent displacement strategy, we have conceived and tested a much simpler strategy, based on incoherent laser pulse injection \cite{Qin2018}{}. Saturating the homodyne detector with external laser pulse indeed presents  several operational advantages over the previous strategy. First, since it is incoherent with the local oscillator, an external laser adds only its own shot noise to the excess noise.  More importantly, relative phase drift compensation is not required for keeping the homodyne in the saturated region. This greatly simplifies Eve's resent setup as shown in Figure \ref{Setup_Ext_Laser}.

In this strategy, saturation is induced by an intense incoherent laser pulse sent along with the resent coherent state. The equivalence of the intensity $I$ of the incoherent laser pulse to the displacement $\Delta$ can be given by $\Delta = \sqrt{\eta_{b}/I_{lo}} (1-2T_{bs})I$, where $\eta_b$ is the efficiency of Bob, $I_{lo}$ is the local oscillator intensity and $T_{bs}$ is the effective transmittance applied to the incoherent laser pulse due to asymmetry of beam splitting ratio and the attenuator (shown in Figure \ref{BalancedHD}). In our case $T_{bs} \approx 49\%$. To bring uniformity in the description of both experimental strategies, since the primary requirement is to induce saturation, we also call  ``displacement'' the effect of this incoherent shift of the measured quadratures.

\begin{figure}[h!]
 \includegraphics[width=0.5\textwidth]{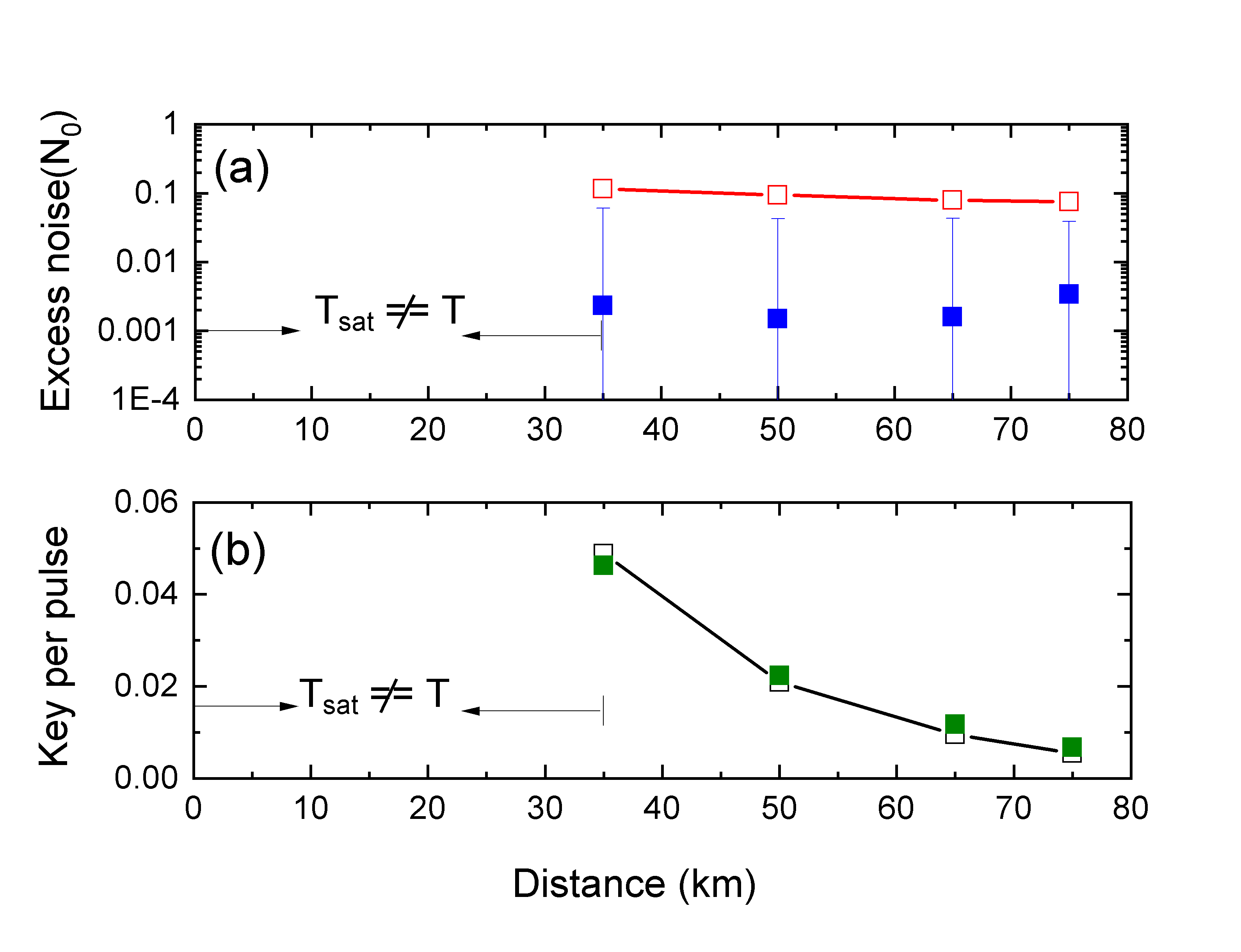}
\caption{Results:- attack with incoherent light. \textbf{(a)} excess noise at Alice. Red squares indicate the null key noise threshold and blue squares the estimated values of $\xi_{sat}$. \textbf{(b)} key rate. Black squares are  simulated values of final key per pulse while Green squares are from the experiment. Error bars are one standard deviation of fluctuations among ten smaller data block of size $10^7$. Success condition of $T_{sat} = T$ can not be fulfilled below 35km.
}           
\label{Result}       
\end{figure}
Since optical phase drift compensation is not needed, saturation attack with incoherent laser pulse can achieve comparatively much performance in terms of quadrature stability and noise, and can  meet
success conditions, provided the channel loss is not too small (low channel loss make it more difficult for Eve to succeed in the intercept-resend attack).
The results in terms of excess noise are given in Figure \ref{Result}(a). The excess noise at Alice has been calculated from the variance of saturated homodyne output experimental data, at various transmission distance between Alice and Bob. It can be seen that  the excess noise is bellow the null key threshold, which indicates Eve's intercept-resend attack remains untraceable. Figure \ref{Result}(b) shows the  maximal value of final key rate per pulse, estimated under collective attacks. Note that the condition $T_{sat} =T$ cannot be met for distance below 35 km - see noise model in section Method. A relaxed attack success condition, where Eve does not maintain $T_{sat} = T$ is given in supplementary information.

\subsubsection*{Rating of the two attacks}

Now that we have defined, and studied the two possible attacks paths to exploiting the saturation vulnerability of the homodyne receiver, we are ready to use Table \ref{Table:Att_pot} to evaluate their Attack Potential. We assume that the hacker Eve tries to obtain as much information as possible about the Target of Evaluation (TOE) design, i.e. we need to assume that Eve has a good knowledge about the specifications of the main components of the QKD system. Part of this information can indeed be found easily online. However, some important details might be system-specific or protected by a non-disclosure agreement between the vendor and the owner of the QKD system. For this reason, for both attacks, the Knowledge factor for the TOE factor is evaluated as \textit{restricted}.

Both attacks rely on the intercept-resend strategy and can in principle be launched in real time.
However, such online implementations of the attacks require to evaluate  the optimal value of the displacement $\Delta$ and of the gain $G$ (see methods): this can be obtained by manually tuning Eve's setup and measure the excess noise due to displacement, as in Figure \ref{NoiseFit}. Assuming a frequent trusted evaluation of the channel loss, this tuning might be quite challenging, especially in the case of the coherent attack, where the tuning precision is inevitably limited by the accuracy of the phase locking. As a result, for  the coherent attack the Windows of Opportunity can be chosen as \textit{difficult}, while \textit{moderate} for the incoherent attack. 
The main differences between the two attack paths are related to the requirements in terms of equipment and expertise.
 As previously explained, the coherent attack requires Eve to resend coherent displaced signal while being successfully phase locked with Alice and Bob. To achieve this, Eve needs to be an \textit{expert} in coherent optical communications, able to control noise at the quantum level and to have access to \textit{bespoke} equipment. On the other hand, the incoherent attack only requires Eve to send an incoherent signal, without worrying about being phase locked with Alice and Bob: this is reflected in a simplified setup (Equimpent \textit{specialized}) and in a lower level of required technical expertise for Eve (Expertise \textit{proficient}).
From Table \ref{Table:Att_pot} we hence obtain an Attack  Potential of 26 and 14 for coherent and incoherent attack respectively. As expected, the coherent attack is rated  as \textit{beyond high}, while the incoherent attack is only rated as \textit{moderate}.


\renewcommand{\arraystretch}{1.4}
\begin{center}
\begin{table*}[htb!]
\centering
\begin{tabular}[t]{  |m{1.9cm}|c|c|c|c|c|c|p{6cm}|  }
\cline{2-8}
\multicolumn{1}{c|}{}  &  \multicolumn{5}{c}{ \cellcolor{lightorange}\textbf{Attack Potential}} & \multicolumn{1}{|c|}{\cellcolor{lightorange} \textbf{Rating}} & \multicolumn{1}{|c}{\cellcolor{lightorange} \textbf{Experimental Results}}\\[3pt]
\hline
\multirow{2}{*}{\cellcolor{lightorange}} & {\footnotesize Exp} &{\footnotesize  KoT} & {\footnotesize  WoO} & {\footnotesize  Equ} & \textbf{AP}& \multirow{2}{*}{\cellcolor{lightcyan}} & \multirow{2}{*}{\begin{tabular}{@{}p{6cm}@{}} {\footnotesize $\checkmark$ Noise model experimentallly characterized} \\ {\footnotesize $\times$ Attack not feasible under noise model} \end{tabular}}\\
\multirow{-2}{*}{\cellcolor{lightorange}\begin{tabular}{@{}c}\textbf{ Coherent}\\[-0.2cm]\textbf{ Attack}\end{tabular}}&{\footnotesize 6}&{\footnotesize 3}&{\footnotesize 10}&{\footnotesize 7}&\textbf{26}&\multirow{-2}{*}{\cellcolor{lightcyan}\begin{tabular}{@{}c@{}} \textbf{Beyond} \\[-0.2cm] \textbf{High} \end{tabular}}&\\[3pt]
\hline 

\multirow{2}{*}{\cellcolor{lightorange}}& {\footnotesize Exp} &{\footnotesize  KoT} & {\footnotesize  WoO} & {\footnotesize  Equ} & \textbf{AP}& \multirow{2}{*}{\cellcolor{orange!30}} & \multirow{2}{*}{ {\footnotesize $\checkmark$ Attack experimentally demonstrated}}\\
\multirow{-2}{*}{\cellcolor{lightorange}\begin{tabular}{@{}c}\textbf{Incoherent}\\[-0.2cm]\textbf{ Attack}\end{tabular}}&{\footnotesize 3} & {\footnotesize 3}&{\footnotesize 4}&{\footnotesize 4}&\textbf{14}&\multirow{-2}{*}{\cellcolor{orange!30} \textbf{Moderate}}&\\
\hline 
\end{tabular}
\caption{Summary of the analysis on the two attacks to the homodyne detection. We have reported the values for each factor of the Attack Potential, namely: Exp. stands for Expertise, KoT for Knowledge of the TOE, WoO for Window of Opportunity and Equ for Equipment. The factors chosen for the analysis are from Common Criteria \cite{CEM}{}.}
\label{V_Assessment}
\end{table*}
\end{center}
\renewcommand{\arraystretch}{1}

\section*{Discussion}

 Quantifying the level of security assurance of a QKD device is a complex undertaking that should be based on a sound and largely recognized methodology. This objective translates into the requirement of standards for QKD security evaluation and certification. Such standards have recently started to be actively developed  within international standardization groups  \cite{QKDStandardISO, QKDStandardETSI} and are now subject to an intense international effort.
 
The practical ability to evaluate QKD security will also require to set up evaluation lab facilities and to train  ``QKD security evaluation engineers'',  able to conduct  penetration testing on QKD systems, both in terms of software and hardware. In that perspective, the experience  accumulated in the context of classical secure hardware, and in particular the use of the Common Criteria methodology by the smartcard industry \cite{SOGIS}{}, is invaluable.

The main message of this article is to point at the relevance of calculating Attack Potential to rate attacks against QKD device, following a methodology already in place to evaluate the security of classical cryptographic hardware. Moreover, Attack Potential can be used as a metric in order to  balance the effort invested at QKD system design stage and at countermeasure development stage to thwart attacks, allowing to prioritize the attacks that constitute the most serious threats in practice. 

 One might however wonder whether this message should be read negatively, from a QKD viewpoint. Does it imply that the security QKD can provide essentially compares to the security that can be reached with classical hardware crypto-systems?

We  want to argue that this not the case, for fundamental reasons: quantum crypto-systems strongly differ from their classical counterparts and provide a security advantage that is not only related to the information-theoretic security versus computational security paradigm.  Quantum cryptography is moreover intrinsically based on models where the inner details of the physical layer are tied to  information-theoretic measurable quantities. For instance, a functional QKD system is by definition sensitive to losses or errors occurring at single photon level. This is in strong contrast with classical systems, where information is typically encoded over a very large number of particles, such as classical optical pulses containing many photons. As a consequence, a classical system is oblivious to leakage occurring at the level of single quanta and cannot match the security level that can be provided, at least in principle, by a quantum crypto-system.

  \begin{figure}[h!]
\includegraphics[width=0.45\textwidth]{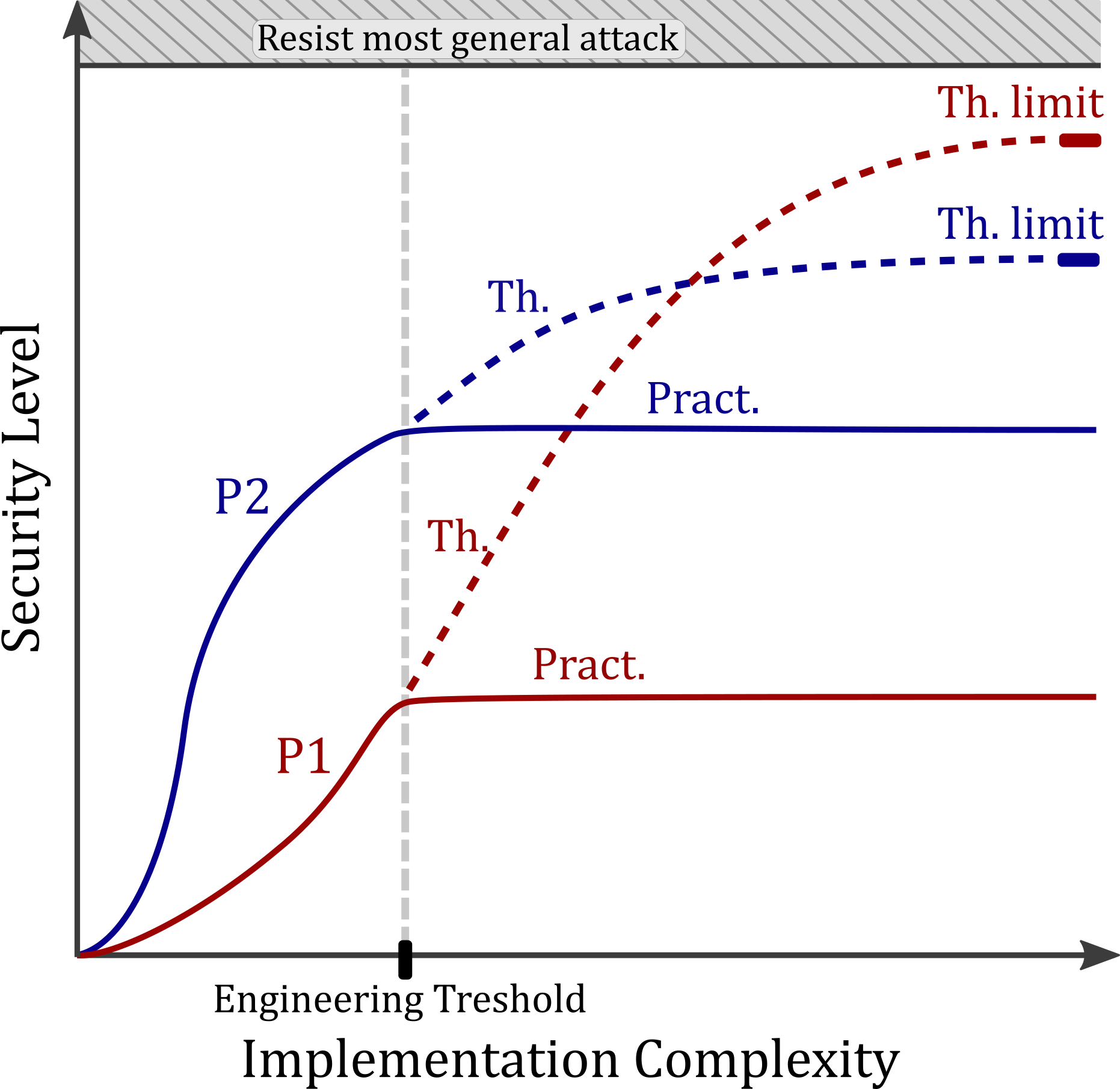}
 \caption{Pictorial representation of the possible divergence between theoretical (th.) and practical (pract.) QKD security. A QKD protocol P1 may have a stronger theoretical security (reachable for a perfect implementation) than another QKD protocol P2. Yet, in practice, QKD protocols can only be operated below a certain implementation complexity level materialized by the engineering threshold, and protocol P2 provides a stronger practical security than protocol P1.}
 \label{risk_analysis}       
\end{figure}

Considering the interplay between QKD implementation complexity and security also leads to an important reassessment. Theoretical security and practical security of a given QKD system  may  indeed significantly  differ, notably when practical security is limited by engineering constraints. This calls to reconsider the absolute security claims sometimes associated with QKD and to adopt a more balanced viewpoint taking implementation complexity into consideration. 
We have depicted on Figure \ref{risk_analysis} an case illustrating this situation: we consider two QKD protocols \cite{Protocol}{}, P1 and P2, where protocol P1  has a more advanced security proof than protocol P2, therefore allowing to claim a higher security level in theory. 
However, it is possible that the protocol P2 has a lower implementation complexity that P1 and that, for the practically reachable implementation complexity corresponding to the engineering threshold, the practical security, i.e. the security that can be reached in practice by the QKD system, is larger with P2 than with P1.

Finally, the use of Attack Potential in QKD  has also implications regarding the security that can be targeted. In particular, optimizing the security level of a given QKD device requires to first thwart attacks with the lowest Attack Potential before focusing on more complex ones. We have moreover explicitly demonstrated, on a live CV-QKD system, how different attacks related to the same theoretical vulnerability - i.e. the non-linear response of the homodyne receiver - can lead to different Attack Potentials.  For a first attack path, detector saturation is reached using a coherent displacement. However, the practicality of this attack is limited due to noise generated from the imperfect phase drift compensation. The second attack path is on the other hand much more dangerous in practice: shining a simple external incoherent laser, it allows to drive the homodyne detector in the non-linear region of its characteristics and to precisely control the excess noise generated from Eve's intercept-resend attack, while meeting the conditions defined for the success of the attack. 

Adapting existing criteria from IT security to the context of quantum cryptography is certainly  a long and challenging path, but it is essential if we aim to make quantum devices relevant in the context of cyber security. We have summarized  in Table \ref{V_Assessment} the results of our security evaluation procedure of two attack paths on CV-QKD, in terms of Attack Potential, illustrating that this methodology can constitute a useful  step towards establishing forward-looking standards for the vulnerability assessment of QKD devices.



\section*{Methods}


\subsubsection*{Common Criteria and Attack Rating}
Common Criteria (CC) is the set of internationally recognized technical standards and configurations for security evaluations of Information Technology (IT) products and technology.  
The terminology and the concepts deployed in the CC aim to be as general as possible. Indeed, they are not intended to restrict the class of IT security problems of which CC is applicable, making them well suited to be extended for quantum communication devices, such as a QKD system.
In simple terms, this comprehensive methodology aims at supporting the needs of three groups with a general interest in evaluation of the security properties of a  certain Target of Evaluation (TOE): owners, developers and evaluators. In particular, what the owner of the TOE  of the device wants is to protect his assets (any possible entity that he places value upon) from possible threat agents, i.e. someone or something that can abuse these assets against the interests of the owner. 

In this context, we seek to offer a standardized framework for evaluating the risks associated to different threats and the effectiveness of the implemented countermeasures for quantum communication devices.
The rating procedure consists in attributing a numeric value to the Attack Potential. In the Common Criteria framework, rating is performed by considering the following factors:
\begin{itemize}
 \setlength\itemsep{0.1em}
\item[a)]Expertise
\item[b)]Knowledge of the TOE
\item[c)]Window of opportunity
\item[d)]Equipment
\item[e)]Elapsed time
\end{itemize}
\begin{center}
\begin{table}[h!]
\centering
\begin{tabular}[t]{  | p{5.5cm}|c|  }

\hline 
\rowcolor{lightorange} \multicolumn{1}{|p{5.5cm}!{\color{lightorange}\vrule}}{\cellcolor{lightorange}} & \\[-0.3cm]
\rowcolor{lightorange} \multicolumn{1}{|p{5.5cm}!{\color{lightorange}\vrule}}{\textbf{Expertise}} & \\
\multicolumn{1}{|p{5.5cm}}{} & \\[-0.43cm]
\hline
& \\[-0.2cm]
Laymen & 0\\
Proficient & 3\\
Expert & 6\\
Multiple experts & 8\\& \\
\hline 
\rowcolor{lightorange} \multicolumn{1}{|p{5.5cm}!{\color{lightorange}\vrule}}{} & \\[-0.3cm]
\rowcolor{lightorange} \multicolumn{1}{|p{5.5cm}!{\color{lightorange}\vrule}}{\textbf{Knowledge of TOE}} & \\
\multicolumn{1}{|p{5.5cm}}{} & \\[-0.43cm]
\hline
& \\[-0.2cm]
Public & 0\\
Restricted & 3\\
Sensitive & 7\\
Critical & 11 \\ & \\
\hline 
\rowcolor{lightorange} \multicolumn{1}{|p{5.5cm}!{\color{lightorange}\vrule}}{} & \\[-0.3cm]
\rowcolor{lightorange} \multicolumn{1}{|p{5.5cm}!{\color{lightorange}\vrule}}{\textbf{Window of Opportunity}} & \\
\multicolumn{1}{|p{5.5cm}}{} & \\[-0.43cm]
\hline
& \\[-0.2cm]
Unnecessary / unlimited access   & 0\\
Easy & 1\\
Moderate & 4\\
Difficult & 10 \\ & \\
\hline 
\rowcolor{lightorange} \multicolumn{1}{|p{5.5cm}!{\color{lightorange}\vrule}}{} & \\[-0.3cm]
\rowcolor{lightorange} \multicolumn{1}{|p{5.5cm}!{\color{lightorange}\vrule}}{\textbf{Equipment}} & \\

\multicolumn{1}{|p{5.5cm}}{} & \\[-0.43cm]
\hline
& \\[-0.2cm]
Standard & 0\\
Specialized & 4\\
Bespoke & 7\\
Multiple bespoke & 9\\ & \\
\hline
\end{tabular}
\caption{Table for the evaluation of the Attack Potential used in the article. Elapsed time factor has not been considered: see text for explanations. For a complete guide on how evaluate those factors refer to the  Common Evaluation Methodology version 3.1 \cite{CEM}{}.}
\label{Table:Att_pot}
\end{table}
\end{center}

Expertise refers to the level of technical expertise required to successfully perform the attack. Clearly an attack that can be mounted by a person with a regular level of education without an advanced knowledge in any specific field should be prioritized. The Knowledge of the TOE involves, instead, the amount of knowledge of the TOE design and operation required: retrieving detailed specifications about the device, for example, might be challenging for an attacker, leading to an higher Attack Potential. Regardless of the information acquired about the TOE, it is possible that, to successfully mount the attack, a previous tuning of the  hacker's setup is needed. This aspect is considered in the Window of Opportunity, together with possible difficulties on getting access to the TOE. 
One last remarkable factor is the level of sophistication of the equipment used in the attack: an attack using equipment easy to obtain and simple to operate is obviously more dangerous than another attack that would require more advanced equipment. Quantum cryptography (QC) even often consider that the attacker Eve might have access to technology not available today such as large quantum computers or long-term quantum memories, and it is clear strength of QC to enable to prove security even in this context. From attack rating viewpoint, such "Quantum equipment" would be considered to have infinite rating, and is such not included in  Table \ref{Table:Att_pot}).
To coherently consider these different factors and evaluate their contribution to the final Attack Potential we assign at each factor one numerical value, following Table \ref{Table:Att_pot}.

In the Common Criteria an additional factor is considered to rate the attacks: the elapsed time, i.e. the time taken to identify a certain vulnerability  and to successfully mount the attack.  To fix a correct timescale, this quantity needs to be compared with the usual time needed for a countermeasure to be applied. 
The vulnerability analysis that we report about in this article, has been performed on a laboratory QKD system.
In this context, the main drivers of the elapsed time such as the product revision lifecycle, or the time during which an attacker could access the QKD system, cannot be meaningfully defined. For these reasons, we did not consider the elapsed time factor factor in our evaluation. We should however state that this factor will become well defined when considering the security of QKD products deployed on real-world networks and should hence be taken into account in future security evaluations of QKD, in conformity with Common Criteria.

In a complete vulnerability analysis, attack rating is sometimes split in two steps, for example in the case of smartcards \cite{SOGIS}{}. The identification step is related to the effort required to create and apply the attack to the TOE for the first time. The exploitation step is then related to the effort required to apply the attack to the TOE knowing the techniques developed in the identification step. Both steps lead to a rating, based on the different factors. For the sake of simplicity, we have not distinguished these two steps in the present article, but we have just followed the general ground rules provided by Common Criteria. This can moreover be justified by the fact that the operational context associated to the exploitation step is essentially missing in the context of laboratory QKD prototype. We hence have rated attacks in a single step, based on the four factors out of five mentioned above (Expertise, Knowledge of the TOE, Window of opportunity and Equipment). This has lead us to adapt the rating methodology and the severity scale presented in Tables \ref{rating}  and \ref{Table:Att_pot}  with respect to the original tables from the Common Evaluation Methodology \cite{CEM}{}.

\subsubsection*{Saturation of balanced homodyne detector}
\begin{figure}[htb!]
\includegraphics[width=0.5\textwidth]{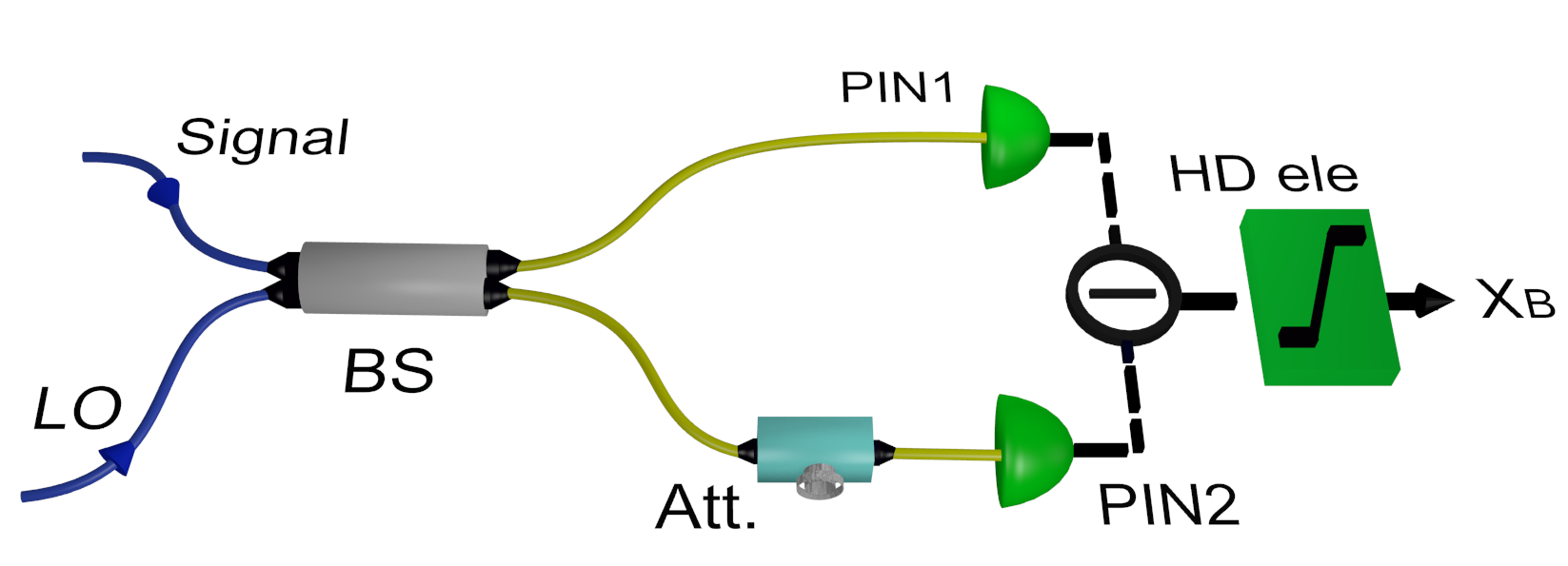}
                \caption{Balanced homodyne detection. An attenuator (Att) in one of the output port of 50/50 beamsplitter (BS) balances  the photo current generated from photodiodes PIN1 and PIN2. Homodyne electronics circuit amplifies the subtracted photo currents. }
                \label{BalancedHD}       
\end{figure}

In a balanced homodyne detection, signal is mixed with intense local oscillator on a 50/50 beam splitter.
Quadrature information is retrieved by subtracting the photocurents generated from two photodiodes  of identical detection efficiency, connected at the output ports of the beamsplitter.  Due to imperfect splitting ratio of beam splitter, as well as efficiency mismatch of photodiodes, it is necessary to add appropriate  attenuation at the respective  output port of the beam splitter, see Figure \ref{BalancedHD}.
 This equalizes the photocurrents and hence sets the mean of the output voltage of the homodyne detection close to zero. This is referred as balancing the homodyne or more precisely ``balancing the homodyne with respect to the local oscillator port". It has been shown that such balancing  is essential to reduce the excess noise due to local oscillator intensity fluctuation \cite{Chi2011}{}.
In case of imperfect balancing one of the photodiodes generates more current than the other. As a result,  the value of the homodyne output  shifts towards the detection limit and this may lead to saturation.  

The reason for saturation is due to the limited amplification factor of homodyne electronic circuitry. In our case circuit is made around Amptek A250 charge amplifier, powered by $\pm$ 5V power supply,  that exhibits detection limit   $\alpha_1$  at -2.5V in the negative DC level and $\alpha_2$  at +3.3V  in the positive DC level (which is observed while interchanging photodiodes). Saturation behavior is also observed while setting low dynamic range   of  data acquisition card (say, $\pm$ 2V) that is used to acquire homodyne output for post processing. In this work, we have set  the data acquisition card range at $\pm$ 5V, thus the linear range is limited solely by the homodyne electronic circuitry.
\subsubsection*{Parameter estimation}

In CV-QKD system that uses GMCS protocol \cite{Grosshans2002}{}, the quantum channel is characterized by its transmission $T$ and its excess noise $\xi$. These parameters are estimated from Alice and Bob modulated and measured quadratures. Under saturation attack, these parameters are modified into $T_{sat}$ and  $\xi_{sat}$.

During the intercept-resend (IR) attack, the quadrature measured by Eve is: $X_M =  X_A + X_0 + X_0^{'}$,
where $X_0$ is  the vacuum noise quadrature due to Alice preparation and $X_0^{'}$ is due to 3dB loss from Eve's heterodyne measurement. The resent signal takes the form:
$X_E = \sqrt{\frac{G}{2}} \left(X_M + X_{N_{A,E}} \right)+\Delta_X +X_0^{''}$. Here, $G$ is the amplification factor to compensate the loss from the heterodyne detection,  $X_{N_{A,E}}$ accounts the technical noise from Alice and Eve, $ X_0^{''}$ is  due to coherent state preparation by Eve. The term  $\Delta_X$  determines the amount of shift  in the mean value of quadrature. The same formalism holds true also for $P$ quadrature.  The parameter estimation takes the form \cite{Qin2016}{}:


\begin{equation}
\begin{array}{l}
T_{sat} = 2 \langle X_A X_{B_{sat}} \rangle^2/ (G\eta_B V_{A}^2)\\[0.3cm]
\xi_{sat} = \frac{2}{G \eta_{_{B}} T_{sat} }( V_{B_{sat}} - G\frac{\eta_B T_{sat}}{2} V_A - N_0 -v_{ele})
\end{array}
\end{equation}

Where, $X_{B_{sat}}$ and $V_{B_{sat}}$  denote quadrature and its variance measured under saturation attack.  One  aspect of the attack worth mentioning  here is that we assume that Eve does not tamper with the shot noise calibration phase.

\subsubsection*{Setup for coherent displacement}

\begin{figure}[htb!]
 \includegraphics[width=0.5\textwidth]{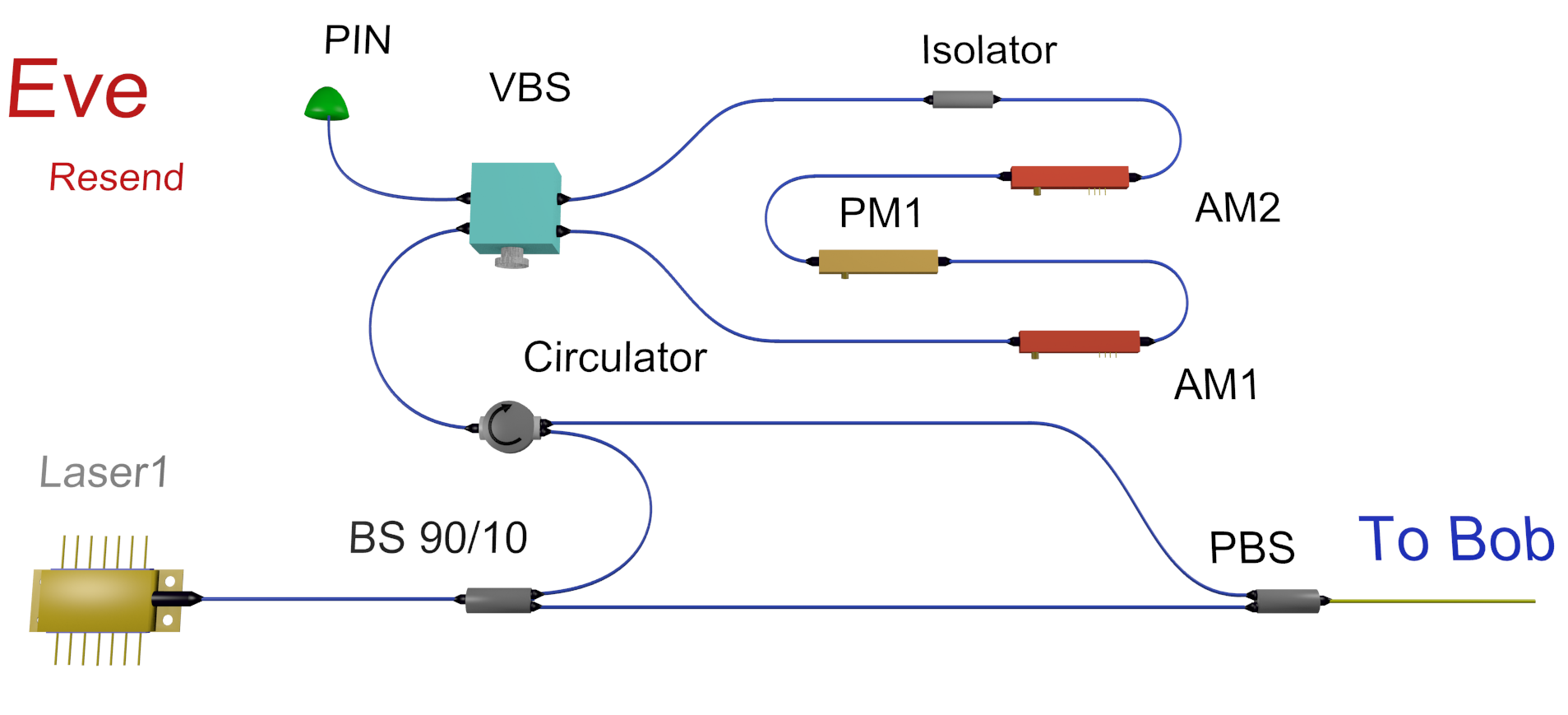}
\caption{Experimental setup for generating displaced coherent state. AM: Amplitude Modulator, PM: Phase Modulator, BS: BeamSplitter. In the Sagnac loop, Gaussian modulated signals are prepared using the AM and PM modulators and are then displaced at the Variable Beam Splitter (VBS), based on a coherent interference between pump. Displaced signals is then sent to Bob along with local oscillator.}
\label{Setup_Displaced}      
\end{figure}

The experimental setup shown in Figure \ref{Setup_Displaced} implements the resend session of the saturation attack. We have implemented  CV-QKD $Eve_{resend}$ system  \cite{Jouguet2013a} using a Sagnac loop realized with variable beamsplitter (VBS).This allows to generate displaced GMCS signals as explained in Results. We have used a 1530.12nm pulsed laser of width 50ns, at  a repetition rate of 1MHz, for generating this displaced signal. Displacing the signal is achieved as follows. The VBS, with splitting ratio $\approx$99.9$\%$, splits the pulse from the circulator into two. Less intense signal pulse in clockwise direction  goes under Gaussian modulation by amplitude modulator (AM1) and phase modulator (PM1) and further heavily attenuated by isolator (connected in reverse  to achieve an attenuation higher than 30dB).  High intense pulse travels along anti-clockwise directions, referred as pump pulse,  meets the signal pulse at VBS and displaces it \cite{Paris1996}{}. The amplitude modulator AM2 controls the intensity of the pump and thence the amount of displacement $\Delta$. A PIN diode attached to the VBS helps to monitor the stability of displacement operation. The Sagnac configuration helps to lock the relative phase of pump pulse and signal pulse to zero. Finally, the circulator directs the displaced signal towards the polarization beam splitter (PBS) that  polarization multiplexes   the local oscillator and displaced signal to the  output fibre channel.

\subsubsection*{Setup for incoherent laser pulse injection}
In this version of attack, Eve sends external laser pulse of 20ns width, along with signal pulse in the same polarization but at different wavelength (1550.12nm). The signal laser  and incoherent laser pulses are synchronized with proper delay.  At Bob station, he performs the same homodyne measurement as in the coherent attack strategy,  where incoherent  laser pulse is polarisation demultiplexed along with signal. It exploits two features of the homodyne setup: imbalance  of the  homodyne  experienced by the light through signal port  and also wavelength dependent splitting ratio of the beam splitter\cite{Huang2014,Ma2013a,Huang2013}{}. By taking into account the wavelength dependent effect and    the attenuator value adjusted for the local oscillator,    the effective transmittance applied to the incoherent laser  is approximately  $T_{bs} \approx 49\%$, while   the   transmittance  applied to local oscillator is about $T_{lo} \approx 50\%\pm 0.05\%$.
Varying intensity of the incoherent light shifts the mean of the homodyne output towards the saturation limit $\alpha_1$ and as a result affects the output variance.

\subsubsection*{Optimizing displacement $\Delta$ and gain $G$}
\begin{figure}[htb!]
 \includegraphics[width=0.5\textwidth]{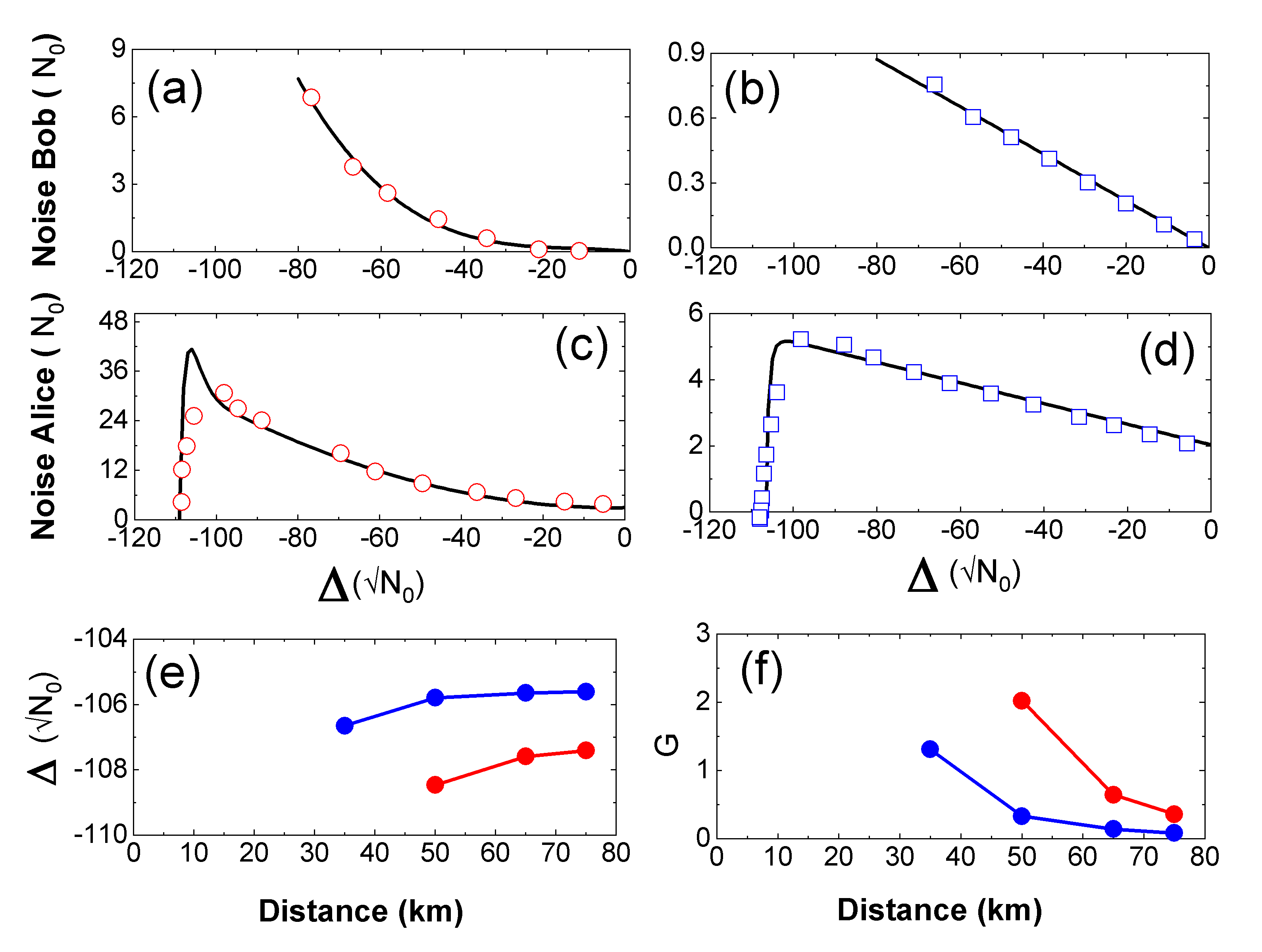}
                \caption{ Excess noise due to displacement\textbf{(a)-(d)}. Red circles and blue squares represents  noise from coherent displacement  and  incoherent light, respectively. Black lines are theoretical fit with respective noise model. (a) and (b) show excess noise at Bob induced by $\Delta$. Noise from coherent displacement  shows quadratic behaviour while  incoherent light adds noise from its own shot noise which is linear. Noise at Alice is shown in (c) and (d).  Optimal values of G (red dots) and   $\Delta$ (blue dots) at various distance, calculated based on noise model from (a) and (b), are shown in  (e) and (f), respectively.}
  \label{NoiseFit}       
\end{figure}
In order to evaluate optimal values of $\Delta$ and $G$ for successful attack,  it is essential to characterize the noises associated with displacement as  well as incoherent laser pulse. In the absence of resent signal, displacement pump/incoherent light is sent to Bob and amount of noise recorded for various values of $\Delta$. This helps to model the excess noise at Bob, shown in Figure \ref{NoiseFit}(a) and (b), and it is taken  into account during optimization of $\Delta$ and $G$.  
The Figure \ref{NoiseFit}(c) and (d) show  excess noise at Alice with respect to $\Delta$. Value of  detection limit $\alpha_1$ is calibrated as $-106 \sqrt{N_0}$ ($-2.5V$ expressed in shot noise unit) for the optimization. For each transmission distance and for respective optimal $V_A$, $\Delta$ and G are calculated such that excess noise falls below the null key threshold. The optimal values are those that correspond to a maximum key rate, with $T_{sat} = T$, shown in Figure \ref{NoiseFit}(e) and (f). It can be seen that at a transmission distance shorter than 50km and 35km, respectively for coherent and incoherent attack strategy,  no values of G and $\Delta$ are able to meet attack success conditions.  In the incoherent attack strategy, the average power of the incoherent light required to reach the detection limit $\alpha_2 = -106\sqrt{N_0}$ is observed as 5.55uW \cite{Qin2013}{}.

\section*{Acknowledgements}
This research was supported by the European Union's Quantum Technology Flagship, through the Horizon 2020 research and innovation programme, under grant agreements CIVIQ No 820466 and OPENQKD No 857156. H.Q  is sponsored by Shanghai Pujiang Program. R. K. acknowledges support from EPSRC grant no EP/T001011/1 (the EPSRC Quantum Communications Hub).   

\section*{Author contributions}
R.K performed the experimental realization of the attacks, F.M performed the attack rating analysis. H.Q  and R.A developed the theoretical models of the attacks.   R.A supervised the project.  All authors contributed to writing the manuscript.
\section*{Additional information}
Supplementary information will be available in the online version of the paper. Correspondence and requests for materials should be addressed to R.K and R.A.
\section*{Competing financial interests}
The authors declare no competing financial interests

\bibliographystyle{naturemag_noURL}
\bibliography{cvref4}

\begin{thebibliography}{10}
\bibitem{rating}
\bibinfo{title}{Our rating scale is slightly different from the usual rating
  scale ({T}able 4 of the {CEM}\cite{CEM}{}). {T}he main reason for that is
  that we had to reshape the boundary values with a factor $4/5$, since in our
  attack potential evaluation we do not consider elapsed time as a factor (see
  {M}ethods). {W}e have also simplified the table a little, by merging
  \textit{basic} and \textit{enhanced-basic}.}
\bibitem{Protocol}
\bibinfo{title}{The term ``protocol'' has to be understood here in a holistic
  sense, ranging from the specifications of the physical implementation of the
  {QKD} devices, to the detail of the logical operations that they perform in
  order to establish a key.}
\end{thebibliography}

\end{document}